\documentclass[superscriptaddress,reprint,amsmath,amssymb,pra]{revtex4-2}
\usepackage{graphicx}
\usepackage{dcolumn}
\usepackage{bm}
\usepackage{booktabs,siunitx}
\usepackage{xurl}
\usepackage{xr}
\begin{document}

\title{Fisher-Based Sensitivity Framework for Rydberg Atom Microwave Electrometry}

\author{Chen-Rong Liu}
\thanks{These authors contributed equally to this work.}
\affiliation{College of Metrology Measurement and Instrument, China Jiliang University, Hangzhou, 310018 China}

\author{Runxia Tao}
\thanks{These authors contributed equally to this work.}
\affiliation{College of Metrology Measurement and Instrument, China Jiliang University, Hangzhou, 310018 China}

\author{Xiang Lv}
\affiliation{Department of Physics, Westlake University, Hangzhou, 310030 China}

\author{Ying Dong}
\affiliation{College of Metrology Measurement and Instrument, China Jiliang University, Hangzhou, 310018 China}

\author{Chuang Li}
\email{chuangli@cjlu.edu.cn}
\affiliation{College of Metrology Measurement and Instrument, China Jiliang University, Hangzhou, 310018 China}

\author {Binbin Wei}
\email{weibb.2009@tsinghua.org.cn}
\affiliation{Qianyuan Laboratory, Xihu District, Hangzhou, Zhejiang Province, 310024 China}

\author{Mingti Zhou}
\email{mtchou@cjlu.edu.cn}
\affiliation{College of Metrology Measurement and Instrument, China Jiliang University, Hangzhou, 310018 China}
\date{\today}

\begin{abstract}
Fisher information provides a rigorous theoretical benchmark for evaluating quantum sensor sensitivity; however, a comprehensive framework for quantifying the fundamental limits of Rydberg-atom microwave electrometers remains lacking. In this work, we establish such a framework by deriving the Fisher information for slope detection and establishing its connection to sensitivity through signal-to-noise ratio, leading to an analytical expression jointly determined by photon shot noise and atomic response. Numerical implementation with real parameters in cesium vapor systems reveals a Fisher-optimized sensitivity below $\mathrm{nV\,cm^{-1}\,Hz^{-1/2}}$, highlighting a substantial potential for sensitivity enhancement in practical experiments through the suppression of technical noise. Importantly, the theory predicts that sub-nanovolt sensitivity is robust against moderate variations in system parameters, thereby delineating both the ultimate sensitivity and optimal operational regime of Rydberg-atom microwave electrometers.
\end{abstract}

\maketitle

\section{\label{intro} Introduction}

Rydberg atoms exhibit exaggerated atomic properties, such as the scaling of dipole-dipole interactions as $n^4$ and the scaling of radiative lifetime as $n^3$ with the principal quantum number $n$, making them a current research hotspot in quantum information, quantum computing, and quantum precision measurement \cite{saffman2010quantum,omran2019generation,browaeys2020many,ebadi2022quantum,zhang2024rydberg,wu2024dependence}.
In recent years, Rydberg-atom microwave electrometers (RAME) have demonstrated remarkable sensitivity by leveraging electromagnetically induced transparency (EIT) and Autler-Townes (AT) splitting phenomena \cite{Adams2019,ASPH2022}. The core detection mechanism relies on measuring the AT splitting interval in EIT spectra, such that $\Delta_{\mathrm{AT}} \propto \Omega_s$ \cite{sedlacek2012,ASPH2022}. This EIT-AT-based approach has led to significant advances \cite{sedlacek2012,sedlacek2013atom,holloway2014broadband,fan2015atom,simons2019rydberg,robinson2021determining,hu2022continuously,bohaichuk2022origin}, achieving sensitivities on the order of $\mathrm{\mu V\,cm^{-1}\,Hz^{-1/2}}$ in room-temperature atomic vapor cells \cite{kumar2017atom,Kumar2017Rydberg}.
A major breakthrough came with the introduction of the atomic superheterodyne detection scheme by Jing \textit{et al.}, which achieved a three-orders-of-magnitude improvement in sensitivity—55 $\mathrm{nV\,cm^{-1}\,Hz^{-1/2}}$ at room temperature \cite{Jing2020}, and 10 $\mathrm{nV\,cm^{-1}\,Hz^{-1/2}}$ at 200 $\mu$K \cite{Tu2024approach}—surpassing traditional EIT-AT protocols. This performance is comparable to that of many-body Rydberg criticality metrology, which has demonstrated sensitivity levels of 49 $\mathrm{nV\,cm^{-1}\,Hz^{-1/2}}$ \cite{Ding2022}, offering yet another promising route for high-precision microwave sensing. Moreover, cavity-enhanced strategies \cite{liu2025cavity,Liang2025cavity} that integrate many-body criticality effects \cite{wang2025high} show potential for pushing sensitivities further into the sub-$\mathrm{nV\,cm^{-1}\,Hz^{-1/2}}$ regime.

The many-body Rydberg criticality and atomic superheterodyne techniques represent paradigm shifts from conventional EIT-AT-based strategies. The former harnesses enhanced susceptibility near quantum critical points, while the latter is based on slope detection \cite{degen2017quantum}. Notably, despite their differing physical principles, both approaches enhance sensitivity through a common mechanism—by utilizing the gradient or slope of spectroscopic signals. In slope detection, the sensor response is linearized by evaluating the derivative $\partial X/\partial \xi$ of a measurable observable $X$ with respect to the parameter of interest $\xi$, at an optimal working point $\xi_0$ \cite{degen2017quantum}. This enables the detection of small variations ($\delta\xi \ll \xi_0$) through the linear approximation $\delta X \approx (\partial X/\partial \xi)\delta\xi$.
The ultimate limit of precision in parameter estimation is governed by the Fisher information $F(\xi)$, as constrained by the Cramér–Rao bound $\mathrm{Var}(\xi) \geq 1/F(\xi)$ \cite{braunstein1994,spehner2013geometric,Paris2009}. 

While Fisher information has been extensively applied in quantum metrology \cite{degen2017quantum}, its systematic adaptation to RAME—especially in optimizing sensitivity—remains underdeveloped. Two theoretical gaps are particularly identified: (i) the absence of a generalized Fisher information formulation specific to slope detection in RAME, and (ii) the lack of a quantitative relationship between conventional sensitivity metrics and $F(\xi)$. Addressing these gaps is further complicated by RAME's indirect measurement architecture, where quantum noise \cite{wu2023quantum} and atomic relaxation dynamics simultaneously affect probe transmission signals, making it challenging to disentangle their respective contributions to sensitivity limitations.

\begin{figure}[ht]
	\centering
	\includegraphics[width=0.95\linewidth]{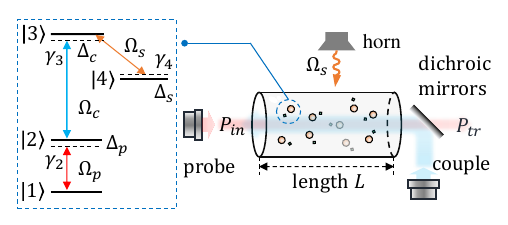}
	\caption{(Color online) \textbf{A schematic of RAME.} Probe laser (input power $P_{\mathrm{in}}$) propagates through a vapor cell (length $L$), with transmitted power $P_{\mathrm{tr}}$. Microwave field (Rabi frequency $\Omega_s$) introduced via horn antenna \cite{sedlacek2012}. Corresponding four-level atomic system configuration is detailed in Table~\ref{tab:table1}.}
	\label{fig:model}
\end{figure}

In this work, we systematically applies Fisher information formalism to slope detection in RAME, establishing a universal framework for determining the theoretical sensitivity benchmark under ideal, shot-noise-limited conditions. Recognizing the pivotal role of signal-to-noise ratio (SNR) in determining measurement sensitivity, we analytically relate SNR to Fisher information via error propagation and parameter estimation theory. Our analysis elucidates how measurement uncertainty arises from quantum fluctuations in the transmitted probe laser power, while atomic relaxation processes govern the overall attenuation of the probe signal (Fig.~\ref{fig:model}). This provides a physical explanation for the sensitivity enhancement observed in atomic superheterodyne detection \cite{Jing2020}. Building upon this general framework, we further perform numerical studies specific to cesium Rydberg-atom systems, identifying critical operating regimes where sensitivity sustains the sub-$\mathrm{nV\,cm^{-1}\,Hz^{-1/2}}$ level, jointly limited by optical shot noise and atomic response. The corresponding experimental configuration, illustrated in Fig.~\ref{fig:model}, employs atomic vapor cell transmission spectroscopy ($P_{\mathrm{in}}\to P_{\mathrm{tr}}$) to detect the microwave-induced linearized power response. 

The remainder of this manuscript is organized as follows: Section~\ref{sec:slope} derives the Fisher information expression and sensitivity limits via error propagation, and formulates general criteria for sensitivity optimization. Section~\ref{sec:numerical} applies the framework to cesium-based Rydberg systems, demonstrating shot-noise-limited performance and evaluating robustness across parameter regimes. A detailed discussion and concluding remarks are provided in Section~\ref{con}. 

\section{\label{sec:slope} Theoretical Framework of Fisher Information and Sensitivity}
In this section, we systematically present our theoretical framework to evaluate the RAME performance. Firstly, we present the basic conversion of microwave-to-optical transduction via slope detection. After that, we derive the Fisher information through error propagation analysis and double check with that via parameter estimates. Finally, we derive the sensitivity limits via signal-to-noise ratio (SNR) of the transmitted probe power, and rigorously connect these two metrics, demonstrating how Fisher information maximization enables identification of optimal operating conditions for RAME. 

\subsection{Slope Detection Principle}
The operational schematic of RAME based on slope detection is illustrated in Fig.~\ref{fig:model}. The core apparatus is an atomic vapor cell, which can be treated as a linear medium \cite{sedlacek2012,steck2024Cesium} at room temperature for the counterpropagating probe and coupling laser. 

The incident ($P_{\mathrm{in}}$) and transmitted ($P_{\mathrm{tr}}$) probe laser powers exhibit a microwave-dependent transmission governed by:
\begin{equation}\label{eq:Ptrans}
	P_{\mathrm{tr}} = P_{\mathrm{in}} \cdot \eta(\Omega_s),
\end{equation}
where $\eta(\Omega_s)$ quantifies the probe attenuation through the atomic medium as a function of the microwave Rabi frequency $\Omega_s$. This attenuation is governed by the Beer-Lambert law:
\begin{equation}\label{eq:eta}
	\eta\left(\Omega_s\right) = \exp\left\{-\frac{2\pi L}{\lambda_{p}}\mathrm{Im}\left[\chi\left(\Omega_s\right)\right]\right\},
\end{equation}
with $\chi(\Omega_s)$ being the complex susceptibility that originates from the microwave-modulated atomic polarization of probe laser transmission \cite{otto2022towards,steck2025quantum}, and $\lambda_p$ the probe laser wavelength. Explicitly, $\chi(\Omega_s)$ depends on the density matrix element ${\rho}_{21}$:
\begin{equation}\label{eq:chi}
	\chi(\Omega_s) = -\frac{2\mathcal{N}_0\mu_{21}^2}{\epsilon_0\hbar\Omega_p}{\rho}_{21},
\end{equation}
where $\mathcal{N}_0$ is the atomic vapor density (m$^{-3}$) calculated via the Clausius-Clapeyron equation \cite{steck2024Cesium}, $\mu_{21}$ denotes the dipole matrix element for the $\vert1\rangle\to\vert2\rangle$ transition (Fig.~\ref{fig:model}), $\Omega_p$ is the probe Rabi frequency, $\epsilon_0$ is the vacuum permittivity, and $\hbar$ is the reduced Planck constant. 

The density matrix element ${\rho}_{21}$ in rotating frame \cite{steck2025quantum} is obtained by solving the quantum master equation under steady-state conditions ($\partial{\rho}/\partial t=0$):
\begin{equation}\label{eq:masEq}
	\frac{\partial{\rho}}{\partial t} = \frac{1}{i\hbar}[{H},{\rho}] + \mathcal{L}{\rho},
\end{equation}
where ${H}$ is the Hamiltonian and ${\rho}$ denotes the density matrix. The term $\mathcal{L}{\rho}$ captures atomic relaxations, including: (1) population decay driven by spontaneous emission \cite{steck2025quantum} and environmental thermal radiation \cite{vsibalic2017arc}, and power broadening due to laser saturation \cite{steck2025quantum} which is incorporated within the decay formalism since it directly modifies effective transition rates; (2) dephasing due to atom-atom collisions \cite{omont1977collisions,Fan2015Microwave,bahrim20013se}, transit-time broadening \cite{sagle1996measurement,fontaine2020}, laser linewidth \cite{Fan2015Microwave,kumar2017atom,finkelstein2023}, and Doppler effect \cite{Fan2015Microwave,finkelstein2023}. Detailed mechanisms of relevant relaxation supporting derivations and analysis are presented in the Supplemental Material\cite{supplemental_material} (see also references \cite{arora2007determination,laser2008laser,Tanasittikosol2011,sobelman2012atomic,rotunno2023inverse,miller2024rydiqule,nagib2025fast}). Solving this master equation yields the probe light transmission $\eta(\Omega_s)$ as the core observable. This quantity directly links the microscopic atomic dynamics, driven by the microwave field $\Omega_s$ to the experimentally measurable macroscopic signal.

Fundamentally, the RAME slope detection mechanism lies on the linear response of the transmitted probe power to small microwave variances. To make this explicit, we consider a perturbations $\delta\Omega_s$ around a reference Rabi frequency $\Omega_0$ (i.e., $\Omega_s = \Omega_0 + \delta\Omega_s$). The linearized power response thus becomes:
\begin{equation}\label{eq:delP}
	\delta P_{\mathrm{tr}} = \left.\frac{\partial P_{\mathrm{tr}}}{\partial \Omega_s}\right\vert_{\Omega_s=\Omega_0}\delta\Omega_s = P_{\mathrm{tr}}\frac{\partial\ln\eta}{\partial \Omega_0}\delta\Omega_s,
\end{equation}
where the second equality follows directly from Eq.~(\ref{eq:Ptrans}). For notational simplicity, we adopt $[\partial/\partial\Omega_s]_{\Omega_s=\Omega_0}=\partial/\partial\Omega_0$ throughout this work. The power response slope is defined as:
\begin{equation}\label{eq:kp}
	\kappa_p \equiv P_{\mathrm{tr}}\frac{\partial\ln\eta}{\partial \Omega_0}.
\end{equation}
The reference Rabi frequency $\Omega_0$ can be optimized to maximize sensitivity. The corresponding microwave electric field perturbation follows:
\begin{equation}\label{eq:EleOmega}
	\delta E_s = \frac{\hbar}{\mu_s}\delta\Omega_s,
\end{equation}
where $\mu_s$ denotes the transition dipole moment between Rydberg states $\vert3\rangle$ and $\vert4\rangle$ \cite{ASPH2022,cai2025polarization}.

This formalism connects linearly the microwave sensing to optical readout: the probe laser induces atomic polarization via $\vert1\rangle\to\vert2\rangle$ transition (Fig.~\ref{fig:model}), generating a susceptibility $\chi(\Omega_s)$ that is parametrically modulated by the microwave field through the Rydberg transition $\vert3\rangle\to\vert4\rangle$. This modulation imprints the microwave Rabi frequency $\Omega_s$ onto the probe transmission $\eta(\Omega_s)$, thereby converting microwave perturbations $\delta\Omega_s$ into measurable optical power variations $\delta P_{\mathrm{tr}}\propto\delta\Omega_s$. The ultimate sensitivity, governed by the transition dipole moment $\mu_s$, and the minimum resolvable $\delta P_{\mathrm{tr}}$, is rigorously quantified through Fisher information as derived in the following subsections.

\subsection{Fisher Information Formalism}
From Eq.~(\ref{eq:Ptrans}), one can conclude that the measurement uncertainty in transmitted probe power $P_{\mathrm{tr}}$ via slope detection stems from optical noise sources dominated by amplitude fluctuations \cite{Venneberg2024Measurement} and the atomic relaxations governing the probe transmission dynamics $\eta(\Omega_s)$. It should be noted that the measured intensity noise spectrum of a laser is inherently insensitive to the laser's phase noise under ideal conditions of linear photodetection, and thus contains no contribution from it \cite{saleh2007fundamentals, yariv2007photonics, gerry2023introductory}. In contrast, amplitude noise can be significantly suppressed to near the shot-noise limit (SNL) through advanced stabilization techniques \cite{kwee2012stabilized, vahlbruch2018, Venneberg2024Measurement}.

Taking SNL as the fundamental noise floor, we can derive the explicit form of the amplitude spectral density through the Wiener-Khinchin theorem and the autocorrelation analysis \cite{danilishin2012quantum, Venneberg2024Measurement}:
\begin{equation}\label{eq:psd}
	S_P = \eta\sqrt{2\hbar\omega_p P_{\mathrm{in}}},
\end{equation}
where $\omega_p$ is the probe optical frequency. This frequency-independent white noise characteristic defines the ultimate precise limit in quantum-correlated laser spectroscopy \cite{Venneberg2024Measurement}. The resultant power fluctuation within the bandwidth $B$ is given by:
\begin{equation}\label{eq:powflu}
	\sigma_{P_{\mathrm{tr}}} = S_P \cdot \sqrt{B} =\eta \sqrt{\frac{\hbar\omega_p P_{\mathrm{in}}}{T}},
\end{equation}
with $B = 1/(2T)$, where $T$ is the integration time. To elucidate how the shot noise propagates through the detection chain, we need to express the incident photon number \cite{gerry2023introductory} over integration time $T$:
\begin{equation}\label{eq:NPin}
	N_{\mathrm{in}} = \frac{P_{\mathrm{in}}T}{\hbar\omega_p},
\end{equation}
representing the total probe photon flux incident onto the atomic vapor cell.

The power fluctuation propagates through the detection chain as quantified by Eq.~(\ref{eq:delP}), ultimately determining the indirect error floor of microwave field measurement:
\begin{equation}\label{eq:errprog}
	\sigma_{\Omega_s} = \frac{\sigma_{P_{\mathrm{tr}}}}{\left|\frac{\partial P_{\mathrm{tr}}}{\partial\Omega_0}\right|} = \left(\sqrt{N_{\mathrm{in}}}\left|\frac{\partial\ln\eta}{\partial \Omega_0}\right|\right)^{-1},
\end{equation}
exhibiting the characteristic $1/\sqrt{N_{\mathrm{in}}}$ scaling. On the other hand, the Cramér-Rao inequality \cite{Paris2009,degen2017quantum} establishes the minimum estimation error:
\begin{equation}\label{eq:CR}
	\Delta\Omega_s \geq \frac1{\sqrt{F(\Omega_s)}},
\end{equation}
where $\Delta\Omega_s = \sqrt{\mathrm{Var}[\Omega_s]}$ denotes the standard deviation. Crucially, this error cannot exceed the quantum noise limit $\sigma_{\Omega_s}$ [Eq.~(\ref{eq:errprog})]:
\begin{equation}
\Delta\Omega_s \geq \sigma_{\Omega_s},
\end{equation}
which reflects the intrinsic constraint of indirect measurement: microwave field uncertainties inherit the optical probe's quantum noise floor.

In this sense, by equating the Cramér-Rao bound [Eq.~(\ref{eq:CR})] with the quantum noise limit [Eq.~(\ref{eq:errprog})], we yields the Fisher information:
\begin{equation}\label{eq:Ferr}
	F(\Omega_s) = N_{\mathrm{in}} \left(\frac{\partial\ln\eta}{\partial \Omega_0}\right)^2,
\end{equation}
which quantifies the maximal achievable precision for microwave field measurement of RAME. To verify the correctness of the above formulation, we also derive the Fisher information using the standard parameter estimation procedure (see Supplemental Material \cite{supplemental_material} for details), which yields the same result.

\subsection{The Limited Sensitivity}
The SNR serves as the fundamental metric for quantifying the sensitivity of RAME. Following quantum metrology conventions \cite{degen2017quantum}, we express the SNR as the ratio of microwave-induced power variation $\delta P_{\mathrm{tr}}$ to intrinsic quantum power fluctuations $\sigma_{P_{\mathrm{tr}}}$:
\begin{equation}
	\mathrm{SNR} = \frac{\delta P_{\mathrm{tr}}}{\sigma_{P_{\mathrm{tr}}}} = \frac{\delta\Omega_s}{\sigma_{\Omega_s}},
\end{equation}
where the SNR of microwave $\Omega_s$ to be measured inherits from the detection chain described in Eq.~(\ref{eq:delP}), including error propagation in Eq.~(\ref{eq:errprog}). The minimum detectable microwave perturbation $\delta\Omega_{s}$, defined at SNR=1 and bounded by the Cramér-Rao inequality \cite{braunstein1994}, is determined through the Fisher information $F(\Omega_s)$ [Eq.~(\ref{eq:Ferr})] as:
\begin{equation}
	[\delta\Omega_{s}]_{\mathrm{min}} = \sigma_{\Omega_s} = \frac{1}{\sqrt{F(\Omega_s)}}.
\end{equation}

Adopt the standard quantum sensing framework with normalized integration time ($T = 1$ s)  \cite{degen2017quantum}, we obtain the fundamental sensitivity limit:
\begin{equation}\label{eq:sens}
\mathcal{E}_s = \frac{\hbar}{\mu_s}[\delta\Omega_{s}]_{\mathrm{min}}\cdot\sqrt{T} = \frac{\hbar}{\mu_s}\sqrt{\frac{T}{F(\Omega_s)}}.
\end{equation}
This expression is intrinsically independent of integration time due to the reciprocal time-scaling relationships in Eqs.~(\ref{eq:NPin}) and (\ref{eq:Ferr}), consistently producing sensitivities in the standard metrological units of $\mathrm{V\,m^{-1}\,Hz^{-1/2}}$. This formulation demonstrates how Fisher information fundamentally governs the ultimate sensitivity limit and dictates its optimization.

The standard quantum limit (SQL) \cite{giovannetti2004quantum} is consequently recovered in our formalism through
\begin{equation}
E_{\mathrm{SQL}} = \frac{\mathcal{E}_s}{\sqrt{T}} = \frac{\hbar}{\mu_{s}}\frac{1}{\sqrt{N_{\mathrm{in}}}\left|{\partial\ln\eta}/{\partial \Omega_{0}}\right|},
\end{equation}
revealing two distinct scaling dependencies: (i) shot noise governed by photon statistics $\propto 1/\sqrt{N_{\mathrm{in}}}$, and (ii) atomic response determined by microwave-modulated transmission variations $\propto 1/\left|{\partial\ln\eta}/{\partial \Omega_{0}}\right|$.

In slope detection, one might naively assume that optimal sensitivity is achieved by simply maximizing the slope $\kappa_p$. However, as shown in Eq.~(\ref{eq:errprog}), the atomic dynamics—characterized by $\eta(\Omega_s)$—also contribute to the noise floor. Therefore, it is crucial to emphasize that maximizing the Fisher information, rather than the slope alone, is essential for determining the optimal experimental parameters to achieve the best sensitivity.

\section{\label{sec:numerical}Numerical Implementation}
With explicit expressions for Fisher information and sensitivity at hand, we can now identify the optimal experimental parameters by maximizing Fisher information.

\subsection{Atomic Relaxations}
In the RAME slope detection scheme, the transmitted probe power depends not only on photon shot noise but also on the atomic response. Therefore, it is crucial to rigorously analyze the various atomic relaxation mechanisms in a quantitative manner. Atomic relaxation mechanisms in RAME can be classified into two distinct categories as previously stated: (i) \emph{decay mechanisms} including spontaneous emission \cite{steck2025quantum}, environmental thermal radiation \cite{vsibalic2017arc}, and laser power broadening \cite{steck2025quantum} that alter population distributions among energy levels; and (ii) \emph{dephasing mechanisms} involving atom-atom collisions \cite{omont1977collisions}, and Doppler effects \cite{Fan2015Microwave,finkelstein2023} that disrupt quantum coherence while preserving population distributions \cite{steck2025quantum}. Experimentally, however, transit-time broadening can be suppressed by enlarging the beam waist radius $w_0$ \cite{sagle1996measurement,fontaine2020,Jing2020}), and the laser linewidth can be reduced to hertz (Hz) or even millihertz (mHz) levels using advanced frequency stabilization techniques \cite{kessler2012sub,dong2015subhertz,huang2017tens,matei20171}. It is therefore justified to adopt an idealized model that neglects transit-time and laser linewidth dephasing. This idealized framework deliberately sets aside the dominant frequency-to-amplitude noise conversion of laser phase noise encountered in practical experiments (analyzed in Supplemental Material \cite{supplemental_material} and Fig.~2). It thereby enables us to isolate and analyze the Fisher information under ideal shot-noise-limited conditions, establishing a clear theoretical benchmark for experimental optimization.

\begin{figure*}[ht]
	\includegraphics{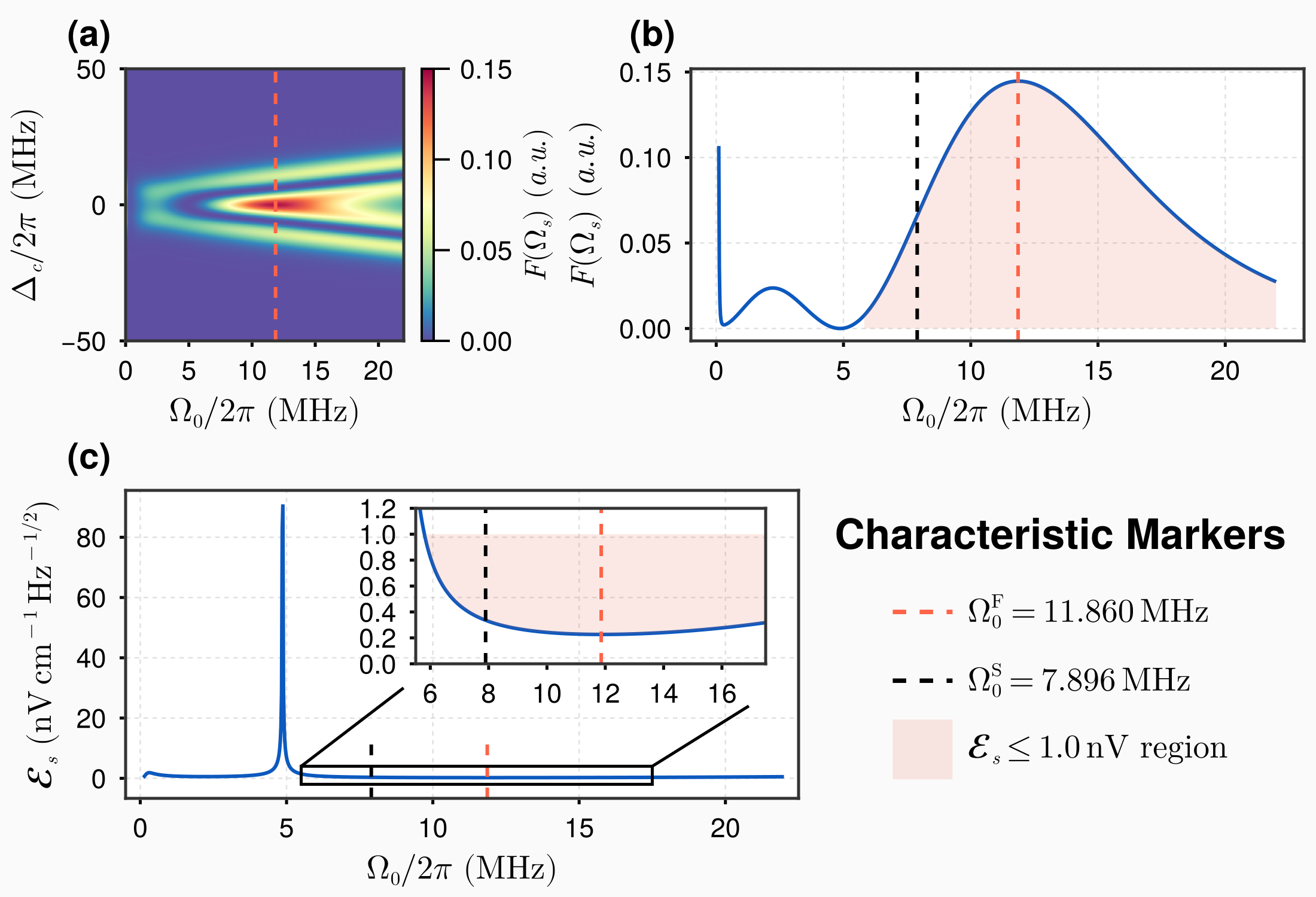}
	\caption{\label{fig2} {\bfseries (Color online) Sub-nV Sensitivity in a Robust Operational Window.} (a) Heatmap of Fisher information $F(\Omega_s)$ [Eq.~(\ref{eq:Ferr})] versus reference microwave Rabi frequency $\Omega_0$ and coupling detuning $\Delta_c$, under $\Delta_p = \Delta_s = 0$. (b) comparison of two characteristic operation points at $\Delta_c = 0$ with vertical lines marking: Fisher information maximum at $\Omega_0^{\mathrm{F}}/2\pi = 11.860\,\mathrm{MHz}$ (dashed orange), and local microwave Rabi frequency $\Omega_0^{\mathrm{S}}/2\pi = 7.896\,\mathrm{MHz}$ (dashed black) in Ref.~\onlinecite{Jing2020}. (c) Microwave field measurement sensitivity $\mathcal{E}_s$ [Eq.~(\ref{eq:sens})] versus $\Omega_0$ at $\Delta_c = 0$ demonstrates a continuous operational regime (shaded region in (b,c) with a wide range of reference microwave Rabi frequency $\Omega_0/2\pi \sim [5,30]\,\mathrm{MHz}$) maintaining shot-noise-limited sub-nV sensitivity ($\mathcal{E}_s \leq 1.0\,\mathrm{nV\,cm^{-1}\,{Hz}^{-1/2}}$), with optimal performance $\mathcal{E}_s^{\mathrm{opt}} = 0.227\,\mathrm{nV\,cm^{-1}\,{Hz}^{-1/2}}$ (dashed orange vertical line). The legend for vertical lines in subplot (c) and its inset (identical to those in panels (b)) is positioned on the right side of (c). All parameters listed in Table~\ref{tab:table1} replicate the experimental parameters from Ref.~\onlinecite{Jing2020}.}
\end{figure*}

Under the rotating-wave approximation (RWA), the rotating-frame Hamiltonian takes the matrix form:
\begin{equation}
	{H}=\frac{\hbar}{2}\begin{bmatrix}0&\Omega_{p}&0&0\\\Omega_{p}&-2\Delta_{p}&\Omega_{c}&0\\0&\Omega_{c}&-2(\Delta_{p}+\Delta_{c})&\Omega_{s}\\0&0&\Omega_{s}&-2(\Delta_{p}+\Delta_{c}-\Delta_{s})\end{bmatrix},
\end{equation}
where $\Omega_{p}$, $\Omega_{c}$, and $\Omega_{s}$ denote probe, coupling, and microwave Rabi frequencies, respectively. The detuning parameters are defined through atomic transition frequencies \cite{Jing2020}:
\begin{eqnarray*}
	\Delta_{p} &=& \omega_{1} + \omega_{p} - \omega_{2}, \\
	\Delta_{c} &=& \omega_{2} + \omega_{c} - \omega_{3}, \\
	\Delta_{s} &=& \omega_{4} + \omega_{s} - \omega_{3},
\end{eqnarray*}
as illustrated in Fig.~\ref{fig:model}. 

For warm atomic vapor, the Lindblad operator describing decay processes is:
\begin{widetext}
	\begin{equation}
		\mathcal{L}_{\rm decay}{\rho} = \left[\begin{array}{cccc}
			\gamma_{2}\rho_{22}+\gamma_{4}\rho_{44} & -\frac{\gamma_{2}}{2}{\rho}_{12} & -\frac{\gamma_{3}}{2}{\rho}_{13} & -\frac{\gamma_{4}}{2}{\rho}_{14} \\
			-\frac{\gamma_{2}}{2}{\rho}_{21} & \gamma_{3}\rho_{33}-\gamma_{2}\rho_{22} & -\frac{\gamma_{2}+\gamma_{3}}{2}{\rho}_{23} & -\frac{\gamma_{2}+\gamma_{4}}{2}{\rho}_{24} \\
			-\frac{\gamma_{3}}{2}{\rho}_{31} & -\frac{\gamma_{3}+\gamma_{2}}{2}{\rho}_{32} & -\gamma_{3}\rho_{33} & -\frac{\gamma_{3}+\gamma_{4}}{2}{\rho}_{34} \\
			-\frac{\gamma_{4}}{2}{\rho}_{41} & -\frac{\gamma_{4}+\gamma_{2}}{2}{\rho}_{42} & -\frac{\gamma_{4}+\gamma_{3}}{2}{\rho}_{43} & -\gamma_{4}\rho_{44}
		\end{array}\right],
	\end{equation}
\end{widetext}
where $\gamma_2$, $\gamma_3$, and $\gamma_4$ denote effective decay rates for transition $\vert2\rangle\to\vert1\rangle$, $\vert3\rangle\to\vert2\rangle$, and $\vert4\rangle\to\vert1\rangle$ respectively, that can be calculated using Alkali Rydberg Calculator \cite{vsibalic2017arc}. 

The dominant collision-induced dephasing primarily arises from Rydberg-ground state atomic collision \cite{omont1977collisions,fabrikant1986interaction,bahrim20013se,weller2019thermal}, incorporating both elastic and inelastic collisions. This dephasing mechanism is described by the Lindblad operator:
\begin{equation}
	\mathcal{L}_{\rm deph}{\rho} = \begin{bmatrix}
		0 & 0 & -\Gamma_{c3}{\rho}_{13} & -\Gamma_{c4}{\rho}_{14} \\
		0 & 0 & 0 & 0 \\
		-\Gamma_{c3}{\rho}_{31} & 0 & 0 & 0 \\
		-\Gamma_{c4}{\rho}_{41} & 0 & 0 & 0
	\end{bmatrix},
\end{equation}
with $\Gamma_{c3}$ and $\Gamma_{c4}$ as collision-induced dephasing rates for the $\vert3\rangle\to\vert1\rangle$ and $\vert4\rangle\to\vert1\rangle$, respectively \cite{weller2019thermal}. 

The total dissipation combines these effects:
\[\mathcal{L}{\rho} = \mathcal{L}_{\rm decay}{\rho} + \mathcal{L}_{\rm deph}{\rho}.\]
Additionally, probe-induced power broadening alters the $\vert2\rangle\to\vert1\rangle$ transition rate via saturated absorption effects \cite{steck2025quantum}, while Doppler broadening is modeled through Boltzmann-weighted averaging of the coherence term ${\rho}_{21}$ \cite{finkelstein2023}. Detailed derivations are provided in the Supplemental Material \cite{supplemental_material}.

\begin{table}[h]
	\caption{\label{tab:table1} Parameters setup for numerical implementation of RAME with $^{133}$Cs atom \cite{Jing2020}.}
	\begin{ruledtabular}
		\begin{tabular}{ll|ll|ll}
			\multicolumn{2}{c|}{\textrm{Laser}} &
			\multicolumn{2}{c|}{\textrm{Atom}} &
			\multicolumn{2}{c}{\textrm{Light-Matter}} \\
			\hline
			$\Omega_{0}$ & 6.947\,$\mathrm{GHz}$ & 
			$\mathcal{N}_0$\footnote{Atom density at $T=298.15\,\mathrm{K}$ (25$^\circ$C), derived from the saturation vapor pressure via the Clausius-Clapeyron equation \cite{steck2024Cesium}} & $4.894\times 10^{16}\,\mathrm{m^{-3}}$ & 
			$L$\footnote{Vapor cell length} & $0.05\,\mathrm{m}$ \\
			
			$\lambda_p$ & $852\,\mathrm{nm}$ & 
			$\mu_{21}$ & $2.5817\,ea_0$\footnote{Transition dipole moments: $\mu_{21}$: $6S_{1/2}(F=4) \to 6P_{3/2}(F^\prime=5)$, $\mu_{32}$: $6P_{3/2} \to 47D_{5/2}$, $\mu_{43}$: $47D_{5/2} \to 48P_{3/2}$ ($1ea_0 = 8.478\times10^{-30}\,\mathrm{C\cdot m}$) \cite{steck2024Cesium}} & $\Omega_{0}/2\pi$ & free paras \\
			
			$\lambda_c$ & $510\,\mathrm{nm}$ & 
			$\mu_{32}$ & $0.0186\,ea_0$ & $\Omega_p/2\pi$ & $9.324\,\mathrm{MHz}$ \\
			
			$P_{c0}$ & $34\,\mathrm{mW}$ & 
			$\mu_{43}$ & $1443.45\,ea_0$ & $\Omega_c/2\pi$ & $0.961\,\mathrm{MHz}$ \\ 
			
			$w_{c0}$ & $1.0\,\mathrm{mm}$ & 
			$\gamma_2/2\pi$ & $5.223\,\mathrm{MHz}$ & $\Delta_L/2\pi$ & $0\,\mathrm{MHz}$ \\
			
			$P_{p0}$\footnote{Probe laser power, and similar for coupling laser.} & $120\,\mu\mathrm{W}$ & 
			$\gamma_3/2\pi$ & $3.982\,\mathrm{kHz}$ & $\Delta_p/2\pi$ & $0\,\mathrm{MHz}$ \\ 
			
			$w_{p0}$\footnote{Probe laser waist radius, and similar for coupling laser.} & $0.85\,\mathrm{mm}$ & 
			$\gamma_4/2\pi$ & $1.745\,\mathrm{kHz}$ & $\Delta_c/2\pi$ & $[-50,50]\,\mathrm{MHz}$ \\
		\end{tabular}
	\end{ruledtabular}
\end{table}

\subsection{Operational Window of Reference Microwave Rabi Frequency}
Our framework enables the systematic optimization of microwave-field measurement sensitivity through Fisher information maximization. This maximization involves locating the zeros of the first derivative of the Fisher information $F_s$ with respect to the reference Rabi frequency $\Omega_{0}$:
\begin{equation}\label{eq:optFerr}
	\frac{\partial F(\Omega_s)}{\partial\Omega_{0}} = 2 N_{\mathrm{in}}\frac{\partial\ln\eta}{\partial\Omega_{0}}\left[-\left(\frac{\partial\ln\eta}{\partial\Omega_{0}}\right)^2+\frac1{\eta}\frac{\partial^2\eta}{\partial\Omega_{0}^2}\right]= 0,
\end{equation}
which identifies parameters ensuring the optimal trade-off between maximized atomic response and suppressed photon-statistical noise, thereby enabling numerical determination of the optimal reference microwave Rabi frequency $\Omega_0^{\mathrm{opt}}$. 

Our optimization strategy fundamentally differs from the slope-based method proposed in Ref.~\cite{Jing2020}. Whereas the latter focuses solely on maximizing the signal slope $\kappa_p = \partial P_{\mathrm{tr}}/\partial\Omega_0$, our Fisher-information-maximization framework explicitly incorporates both the linear response to the external signal and the quantum measurement noise. This distinction is critical: since the signal-to-noise ratio (SNR) is determined by both the signal slope and the associated noise, optimizing the slope alone fails to capture the full picture. Consequently, our approach provides a rigorous and universally valid route to identifying optimal sensitivity limits.

Figure.~\ref{fig2} compares our approach with that of Ref.~\onlinecite{Jing2020}, using identical experimental parameters as listed in Table \ref{tab:table1}. Panel (a) shows the Fisher information $F(\Omega_s)$ under resonant conditions ($\Delta_s = \Delta_p = 0$), calculated with $\Omega_p^{\mathrm{S}}/2\pi = 9.324\,\mathrm{MHz}$ and $\Omega_c^{\mathrm{S}}/2\pi = 0.961\,\mathrm{MHz}$. The condition $\Delta_c = 0$ ensures three-photon resonance, consistent with the experimental setup in Ref.~\onlinecite{Jing2020}. As shown by the dashed vertical lines in panels (b) and (c), the reference microwave Rabi frequency $\Omega_0^{\mathrm{S}}/2\pi = 7.896\,\mathrm{MHz}$ used in Ref.~\onlinecite{Jing2020} differs from the Fisher-optimal value $\Omega_0^{\mathrm{F}}/2\pi = 11.860\,\mathrm{MHz}$. A quantitative comparison in panel (c) reveals that the Fisher-optimized protocol achieves a superior limited sensitivity of $\mathcal{E}_s = 0.227\,\mathrm{nV\,cm^{-1}\,Hz^{-1/2}}$. Notably, a broad operational window $\Omega_0/2\pi \sim [5, 30]\,\mathrm{MHz}$ also yields sensitivities below $1.0\,\mathrm{nV\,cm^{-1}\,Hz^{-1/2}}$, underscoring the enhanced robustness and tunability of the optimized approach for practical implementation.

The existence of this broad operational window reveals a key physical insight: the Rydberg sensor possesses an intrinsic fault tolerance against fluctuations in the reference microwave field. This is a critical practical advantage, as it demonstrates that achieving and maintaining optimal sensitivity does not require exquisite stabilization of the microwave power—a parameter that can be challenging to control with high precision in practical settings. This robustness lowers the technical barrier for implementing high-performance RAME systems. 

Beyond this robustness in parameter selection, the theoretical predictions based on the current parameter configuration yield a sensitivity limit of $0.227\,\mathrm{nV\,cm^{-1}\,Hz^{-1/2}}$, which significantly outperforms existing experimental records—$55\,\mathrm{nV\,cm^{-1}\,Hz^{-1/2}}$ with vapor cells \cite{Jing2020} and $10\,\mathrm{nV\,cm^{-1}\,Hz^{-1/2}}$ with cold atoms \cite{Tu2024approach}. The substantial gap between the experimental records and the shot-noise-limited benchmark underscores that the primary path to improved performance lies in suppressing technical noise to approach its predicted fundamental limit. This conclusion is corroborated by both the dominance of laser frequency noise in the optical readout chain (accounting for $>$ 99\% of noise above 100 kHz) \cite{Jing2020} and our direct numerical analysis of its impact (see Supplemental Material \cite{supplemental_material}, Fig.~2), which together pinpoint technical noise suppression as the critical challenge.

\begin{figure*}[ht]
	\includegraphics{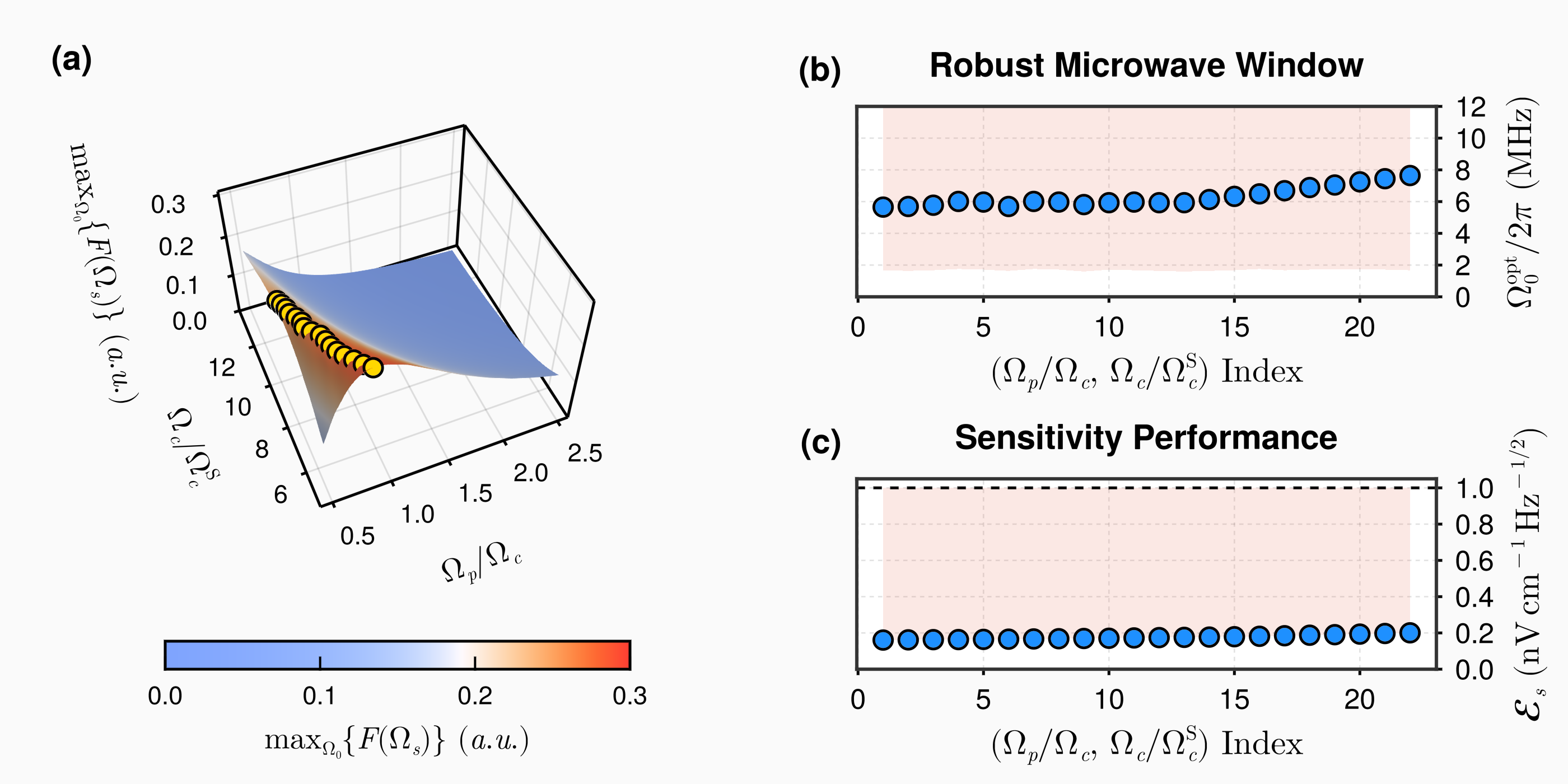}
	\caption{\label{fig3} {\bfseries (Color online) Universal Optimization of Microwave Sensing Parameters.} 
	(a) 3D surface of Fisher information $\mathrm{max}_{\Omega_0}\{F(\Omega_s)\}$ maximized over reference microwave $\Omega_0$, mapped as functions of probe-to-coupling ratio $\Omega_p/\Omega_c$ and normalized coupling strength $\Omega_c/\Omega_c^{\mathrm{S}}$ under three-photon resonance ($\Delta_p=\Delta_s=\Delta_c = 0$). Gold circle marks local maxima along the ridge suggests optimal $(\Omega_p,\Omega_c,\Omega_0)$ configurations. (b,c) Parameter-optimized characteristics along decline the ridge in (a): optimal reference microwave Rabi frequency $\Omega_{0}^{\mathrm{opt}}$ and corresponding sensitivity $\mathcal{E}_s^{\mathrm{opt}}$, respectively. Shading band indicates robust operational window sustaining sub-nV sensitivity $\mathcal{E}_s^{\mathrm{opt}} \leq \mathcal{E}_s \leq 1.0\,\mathrm{nV\,cm^{-1}\,{Hz}^{-1/2}}$ as in Fig.~\ref{fig2}(b,c). All configurations along the ridge in (a) achieve $\mathcal{E}_s^{\mathrm{opt}} <1.0\,\mathrm{nV\,cm^{-1}\,Hz^{-1/2}}$.}
\end{figure*}

\subsection{Parameter Space Robustness and Optimization Landscape}
The observed parametric resilience motivates systematic characterization of broader parameter spaces. Figure~\ref{fig3} analyzes robustness through Fisher-information-optimized sensitivity. 

Panel (a) shows a 3D surface plot of maximum Fisher information over $\Omega_0$ as functions of probe-to-coupling ratio $\Omega_p/\Omega_c$ and normalized coupling strength $\Omega_c/\Omega_c^{\mathrm{S}}$ under three-photon resonance ($\Delta_p=\Delta_s=\Delta_c = 0$). Local maxima along the ridge are marked with gold circles. Panels (b) and (c) reveal parametric stability along this ridge: both the optimal $\Omega_{0}^{\mathrm{opt}}$  and the corresponding $\mathcal{E}_s^{\mathrm{opt}}$ exhibit approximate constancy across a wide range of laser setup $\Omega_p,\Omega_c$. Notably, deviations of $\pm75$\% in $\Omega_0$ (shaded pink band) induce sensitivity variations $\leq 0.8\,\mathrm{nV\,cm^{-1}\,Hz^{-1/2}}$, while the shaded region maintains $\mathcal{E}_s^{\mathrm{opt}} \leq \mathcal{E}_s \leq 1.0\,\mathrm{nV\,cm^{-1}\,{Hz}^{-1/2}}$. 
This robust maintenance of optimal sensitivity across a wide parameter space demonstrates that the Fisher-information-based optimization constitutes a generalizable design principle, rather than a procedure reliant on a specific, fragile set of laser parameters. Consequently, this implies that the framework can reliably guide the configuration of RAME systems under diverse experimental conditions, accommodating variations in laser powers that inevitably occur between different setups or over time. The path to optimal sensitivity it delineates is not only superior but also experimentally accessible and reproducible.

\section{\label{con} Conclusion}
The application of a Fisher-information framework enables a rigorous determination of the fundamental sensitivity limits in RAME under slope detection protocols. By integrating parameter estimation theory with error propagation analysis, we identify Fisher information as the key metric governing measurement precision and explicitly relate it to the signal-to-noise ratio of the transmitted probe light. Our analysis reveals that the ultimate sensitivity is constrained by two primary factors: (1) optical shot noise, scaling as $1/\sqrt{N_{\mathrm{in}}}$, arising from photon statistics, and (2) microwave-encoded atomic nonlinear response in the probe transmission, captured by the derivative $\partial\ln\eta/\partial\Omega_0$. Crucially, we clarify that the noise floor is determined by optical intensity detection of the transmitted probe power—rather than direct atomic state projection—with the fundamental limit set by the quantum statistics of light (shot noise), despite of both noise types being formally consistent with the standard quantum limit.

Therefore, it is crucial to distinguish the purpose of our framework from how it should be interpreted. While our analysis provides the fundamental sensitivity limit under the stated idealizations, any practical system will be affected by additional technical noise. The large discrepancy between our prediction and experimental results does not stem from a failure of the model but rather validates its utility as a benchmark: it clearly demonstrates that the primary path to bridging this gap lies in the meticulous suppression of classical noise sources, most notably laser frequency noise, to approach the shot-noise-limited regime.

To validate our framework, numerical implementation using cesium-133 parameters from Ref.~\onlinecite{Jing2020} predict a sensitivity limit of $0.227\,\mathrm{nV\,cm^{-1}\,Hz^{-1/2}}$ , substantially surpassing existing experimental records ($55\,\mathrm{nV\,cm^{-1}\,Hz^{-1/2}}$), quantifying the substantial potential for enhancement available through the suppression of technical noise. Additionally, our numerical results reveal robustness near the optimal operating point: significant deviations in probe-to-coupling power ratios ($\Omega_p/\Omega_c$) and normalized coupling strength $\Omega_c/\Omega^{\mathrm{S}}_c$ result in only minimal sensitivity degradation. This intrinsic tolerance substantially relaxes experimental implementation constraints, making sub-nanovolt sensitivity achievable under realistic conditions.

\begin{acknowledgments}
This work was supported by the National Natural Science Foundation of China (Grant NO. 12304545 and 12475042) and the National Key Research and Development Program of China (Grant NO. 2023YFF0718400). The authors acknowledge the use of AI-powered language editing tools to enhance the readability and grammatical accuracy of this manuscript while preserving scientific rigor and technical terminology.
\end{acknowledgments}

\section*{Data Availability Statement} 
The data that support the findings of this article are openly
available~\cite{data_availability}.

\bibliography{RydSensFishRefs}

\end{document}